\documentclass[preprint2]{aastex62}

\usepackage{mathtools}

\graphicspath{{./}{figures/}}
\usepackage{lineno}

\shorttitle{\textcolor{black}{Investigating the Dainotti Relation in GRB through Multipolar EM Radiation}}
\shortauthors{}

\begin{document}

\title{\textcolor{black}{Investigating the Dainotti Relation in Gamma‐Ray Bursts through Multipolar Electromagnetic Radiation} }

\correspondingauthor{ Rahim Moradi; Yu Wang}

\author{ Emre S. Yorgancioglu}
\affil{State Key Laboratory of Particle Astrophysics, Institute of High Energy Physics, Chinese Academy of Sciences, Beijing 100049, China.}
\affil{University of Chinese Academy of Sciences, Chinese Academy of Sciences, Beijing 100049, China}

\author{Daban Mohammed Saeed}
\affil{State Key Laboratory of Particle Astrophysics, Institute of High Energy Physics, Chinese Academy of Sciences, Beijing 100049, China.}
\affil{University of Chinese Academy of Sciences, Chinese Academy of Sciences, Beijing 100049, China}

\author{Rahim Moradi}
\email{rmoradi@ihep.ac.cn}
\affil{State Key Laboratory of Particle Astrophysics, Institute of High Energy Physics, Chinese Academy of Sciences, Beijing 100049, China.}

\author{Yu Wang}
\email{yu.wang@inaf.it}
\affil{ICRA and Dipartimento di Fisica, Sapienza Universit\`a di Roma, P.le Aldo Moro 5, 00185 Rome, Italy.}
\affil{ICRANet, P.zza della Repubblica 10, 65122 Pescara, Italy.}
\affil{INAF -- Osservatorio Astronomico d'Abruzzo, Via M. Maggini snc, I-64100, Teramo, Italy.}

\begin{abstract}

The Dainotti relation empirically connects the isotropic plateau luminosity ($L_X$) in gamma-ray bursts (GRBs) X-ray afterglows to the rest-frame time at which the plateau ends ($T_a^*$), enabling both the standardization of GRBs and their use as cosmological probes. However, the precise physical mechanisms underlying this correlation remain an active area of research. Although magnetars, highly magnetized neutron stars, have been proposed as central engines powering GRB afterglows, traditional dipole spin-down radiation models fail to account for the full diversity of observed behaviors. This limitation necessitates a more comprehensive framework. We propose that multipolar magnetic field emissions from magnetars offer a \textcolor{black}{plausible} explanation for the Dainotti relation. Unlike simple dipole fields, higher-order multipolar configurations enable more complex energy dissipation processes. The coexistence of multiple components can \textcolor{black}{plausibly} explain the range of afterglow decay indices found from a sample of 238 GRBs with plateau features from the Swift-XRT database up to the end of December 2024, the majority of which deviate from the dipolar prediction of $\alpha = -2$, and more crucially, the spin-down physics yields a link between $L_X$ and $T_a^*$  in a way that preserves the Dainotti correlation with a slope of $b = - 1$, independent of the specific multipole order. Moreover, we find that the inclusion of higher order multipoles \textcolor{black}{can} explain the range of plateau energies found in the Dainotti relations. Thus, a unified picture emerges in which multipolar fields \textcolor{black}{are able to} reproduce both the slope and the normalization of the correlation.

\end{abstract}


\section{Introduction} 
\label{sec:intro}

Gamma-ray bursts (GRBs) are among the most energetic phenomena in the universe, characterized by intense bursts of gamma radiation followed by longer-lasting afterglows across multiple wavelengths \citep{2004IJMPA..19.2385Z,2004RvMP...76.1143P, 2006RPPh...69.2259M, 2014ARA&A..52...43B, 2015PhR...561....1K, 2015JHEAp...7...73D}. Over the past decades, observational studies have uncovered numerous correlations in GRB properties that provide valuable insights into their underlying physics; see e.g., \cite{2012A&A...538A.134X, 2018pgrb.book.....Z, 2018AdAst2018E...1D} and references therein. One such correlation, the Dainotti relation, links the isotropic plateau luminosity $L_X(T_a) = L_X$ and the rest-frame time at which the plateau ends in the X-ray afterglows of GRBs $T_a^*$, with the relation $\log L_X = \log a + b \log T_a^*$, where the intrinsic slope is $b = -1.07_{-0.14}^{+0.09}$ \citep{2008MNRAS.391L..79D, 2010ApJ...722L.215D, 2011ApJ...730..135D, 2011MNRAS.418.2202D, 2013ApJ...774..157D}. 
 This empirical relation has attracted significant attention due to its potential to shed light on the physical mechanisms of the central engine, as well as serving as a cosmological tool \citep{Cao:2022yvi,2022MNRAS.510.2928C, Dainotti:2022ked, Favale:2024lgp}.

 Several mechanisms have been proposed to explain the Dainotti correlation, including fallback accretion onto a black hole \citep{2009ApJ...700.1047C, 2011ApJ...734...35C}, but magnetars have emerged as particularly promising candidates \citep{2001ApJ...552L..35Z, 2010MNRAS.402..705L, 2011A&A...526A.121D, 2011AIPC.1358..195R, 2012IAUS..279..297O, 2012Sci...338.1445N, 2013MNRAS.430.1061R, 2014ApJ...785...74L, 2014MNRAS.443.1779R, 2015ApJ...805...89L, stratta2018magnetar}. The spin-down of a newly born magnetar powers the radiation that can produce the GRB afterglow, potentially accounting for both the observed plateau phase and its subsequent power-law decay \citep{BERNARDINI201564}. Under the assumption of purely dipolar spin-down, the resulting model correctly reproduces a Dainotti correlation slope of $-1$. However, it also yields a post-plateau decay index of $-2$, which conflicts with most Swift-XRT observations, where decay indices typically lie between $-1$ and $-2$ (see Figure \ref{fig:plateau_hist} and \cite{Wang:2024fun}).

 Recent studies address this discrepancy by incorporating higher-order multipolar magnetic fields, extending beyond the conventional dipole approximation \citep{10.1093/mnras/stt872,2017A&A...600A..98D,2020ApJ...893..148R,Wang:2024fun}. Stronger localized fields associated with higher-order moments significantly influence the magnetar’s spin-down behavior and resulting radiation, thereby accommodating the observed range of decay indices. Building on these insights, we propose that higher-order magnetic moments can provide a \textcolor{black}{plausible} explanation for the Dainotti relation. The contribution from multipoles \textcolor{black}{can} reproduce the correlation between plateau luminosity and duration with a slope of $b \sim -1$, regardless of the multipole order, while also allowing for the possibility of x-ray afterglow lower decay indices.

\textcolor{black}{ This paper is organized as follows. In Section~\ref{sec:grb} we summarize the formation channels of magnetars in both long and short GRBs and review the canonical dipole‐driven energy‐injection model. In Section~\ref{sec:model} we introduce our generalized multipolar spin‐down formalism, deriving the luminosity and torque contributions of each harmonic order. In Section~\ref{sec:general_multipole} we develop the analytic luminosity evolution and discuss vacuum versus wind spin‐down regimes. In Section~\ref{sec:Dai} we demonstrate how multipolar spin‐down  reproduces the slope and normalization of the Dainotti relation and compare to the Swift‑XRT sample. We assess the impact of radiative efficiency and jet geometry on the inferred correlation in Section~\ref{sec:eff}. Finally, in Section~\ref{sec:conclusion} we discuss the implications of our results, outline key caveats, and present our conclusions.}

\section{Magnetar Formation and GRBs}\label{sec:grb}

Magnetars formed via binary neutron star (NS) mergers or single-star collapse play a critical role in GRBs. Long-duration GRBs (LGRBs) are often linked to type Ic supernovae, leaving a newborn NS or magnetar at the explosion's core, while short-duration GRBs (SGRBs) are thought to arise from compact binary mergers \citep{2012grb..book..169H,2018pgrb.book.....Z}.

Around half of GRB X-ray afterglows exhibit a canonical decay pattern, including a plateau phase, which is often explained by the magnetar central engine hypothesis \citep{2001ApJ...552L..35Z, BERNARDINI201564}. A rapidly spinning magnetar dissipates rotational energy, providing a plausible mechanism for the observed X-ray plateaus and flares \citep{2006ApJ...642..354Z, 2006ApJ...642..389N,2011MNRAS.413.2031M, Wang:2024fun, Hashemi:2024nmh}. Further evidence of this mechanism emerges from a comprehensive exploration of the Swift database, which reveals numerous GRBs with similar temporal characteristics.

The UK Swift Science Data Centre\footnote{\href{https://www.swift.ac.uk/xrt_products/index.php}{$\rm https://www.swift.ac.uk/xrt\_products/index.php$}}, which updates the dataset initially compiled by \cite{2009MNRAS.397.1177E}, provides properties of GRBs, including spectra, positions, light curves, and temporal decay indices. Our analysis, which updates the results presented in \cite{Wang:2024fun}, originally identified 204 GRBs (as of March 2021) with observed plateau structures in their afterglows, followed by a decay phase well-described by a power-law function, $\propto t^{\alpha}$, where $\alpha$ represents the decay index. We have now expanded the sample to include an additional 34 GRBs with the same characteristics observed up to December 2024, bringing the total to 238 GRBs.

\begin{figure}
\centering
\includegraphics[width=1\hsize,clip,angle=0]{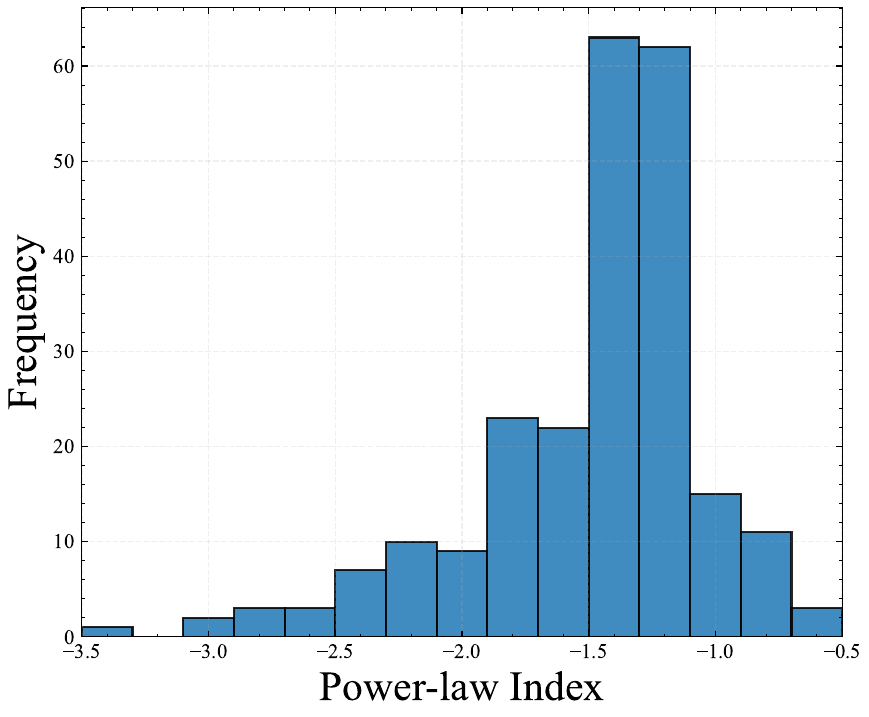}
\caption{A total of 238 GRBs exhibiting a plateau structure were recorded by Swift-XRT up to the end of December 2024. The histogram displays the power-law decay indices of this sample. The majority of these indices (79\%) fall between \ensuremath{-2} and \ensuremath{-1}, with the median and mean values being \ensuremath{-1.39} and \ensuremath{-1.54}, respectively.}
\label{fig:plateau_hist}
\end{figure}

Figure~\ref{fig:plateau_hist} presents the distribution of the power-law decay indices, which mainly range between $-1$ and $-2$, with median and mean values of $-1.39$ and $-1.54$, respectively. Although traditional models emphasize the dominance of dipole radiation $(\alpha = -2)$, the inclusion of higher-order magnetic field components is essential for capturing these observed decay profiles, as was shown by \citep{Wang:2024fun}. The characteristic plateau phase, followed by a steep decline, aligns with predictions from models incorporating multipolar magnetic fields, which extend beyond the simplistic dipolar framework, reinforcing the role of magnetar multipolar radiation as a key mechanism driving GRB afterglows and also highlighting the importance of complex magnetic field structures in shaping observed behaviors \citep{10.1093/mnras/stt872,2017A&A...600A..98D,2020ApJ...893..148R,Wang:2024fun}. 

We also notice in Figure~\ref{fig:plateau_hist}, $\sim 5\%$ of GRBs exhibit a decay shallower than $-1.0$, this is likely due to the inadequacy of late-time data, which introduces large uncertainties in the decay index, and a few samples with late-time flares can flatten the index. Most GRBs with decay indices outside the -1 to -2 range show steeper decays ($\sim 15 \%$ of total samples), which are most plausibly explained by jet breaks, as the relativistic jet decelerates and its beaming angle becomes visible, the afterglow flux drops sharply. 

\textcolor{black}{It is important to note that attributing the post-plateau decay slopes to magnetar spin-down relies on the assumption that the observed X-ray light curves are powered by the internal dissipation of the magnetar wind with constant radiative efficiency and geometric parameters. However, the most common interpretation of the X-ray plateau emission in GRBs is that it arises from synchrotron radiation of the external shock, as described by the standard afterglow model \citep{1998ApJ...497L..17S,2001ApJ...552L..35Z}. In this framework, the plateau phase can result from continuous energy injection into the external shock, possibly due to prolonged central engine activity or variations in the circumburst medium \citep{2006ApJ...642..389N,2006ApJ...642..354Z,2003ApJ...591.1086G}. {\em If the plateau arises from standard external‐shock synchrotron (e.g.\ continuous energy injection), our multipolar spin‐down interpretation no longer applies.}}

\textcolor{black}{However, several studies have reported X-ray plateaus that challenge the standard afterglow model, suggesting the need for additional energy injection mechanisms. For instance, the X-ray afterglow of GRB 070110 exhibited a steep decay $\sim 20$ ks after the burst, ending an apparent plateau, which was interpreted as being powered by a continuously active central engine rather than an external shock \citep{2007ApJ...665..599T}. Similarly, the X-ray afterglow of GRB 130831A showed an ``internal plateau'' with a decay slope of $\sim 0.8$, followed by a steep drop, which was modeled using a magnetar central engine model \citep{10.1093/mnras/stv2280}.}

\section{ Multipolar Spin-down}\label{sec:model}

Magnetars exhibit complex magnetic fields, including dipolar and higher-order multipolar components, spanning strengths from below the quantum critical field $B_Q \approx 4.4 \times 10^{13}~\mathrm{G}$ to $\sim 10^{16}~\mathrm{G}$ \citep{1992ApJ...395..250G, 2002ApJ...574..332T, 2010Sci...330..944R, 2014ApJ...781L..17R,  2016PNAS..113.3944G, 2020ApJ...905L..31G}. Multipolar fields dominate early-time spin-down luminosity and GRB afterglows, complementing the long-term effects of dipolar fields \citep{1999A&A...345..847G, Wang:2024fun}.

Building on our previous work \citep{Wang:2024fun}, we adopt a simplified framework to isolate the role of multipolar magnetic fields, assuming a vacuum environment and modeling the NS as a point source, while neglecting its finite size and temporal field evolution. Although toroidal fields affect spin down via internal dynamics and environmental interactions \citep{2018ApJ...852...21G, 2021Univ....7..351I}, we omit them for simplicity.

In more complex scenarios, magnetic field decay and toroidal components driven by Ohmic dissipation, Hall drift, and ambipolar diffusion substantially weaken the magnetic field strength and its associated radiation \citep{1992ApJ...395..250G, 2006RPPh...69.2631H, 2024MNRAS.528.5178S}. Plasma effects further reduce the field strength by up to 1.5 times and spin-down luminosity by a factor of two relative to vacuum cases \citep{2006ApJ...648L..51S}. These processes lower the braking index from 3 (dipolar) to as low as 2.5, shifting the power-law index for luminosity decay from $-2$ to $-2.33$. Such steeper decay curves suggest that systems deviating from dipolar behavior likely involve significant contributions from higher-order multipolar components.

\section{Generalized Multipolar Spin-Down and Luminosity Formalism}
\label{sec:general_multipole}

\subsection{Electromagnetic Angular Velocity Evolution}

\begin{figure}
\centering
\includegraphics[width=1\hsize,clip,angle=0]{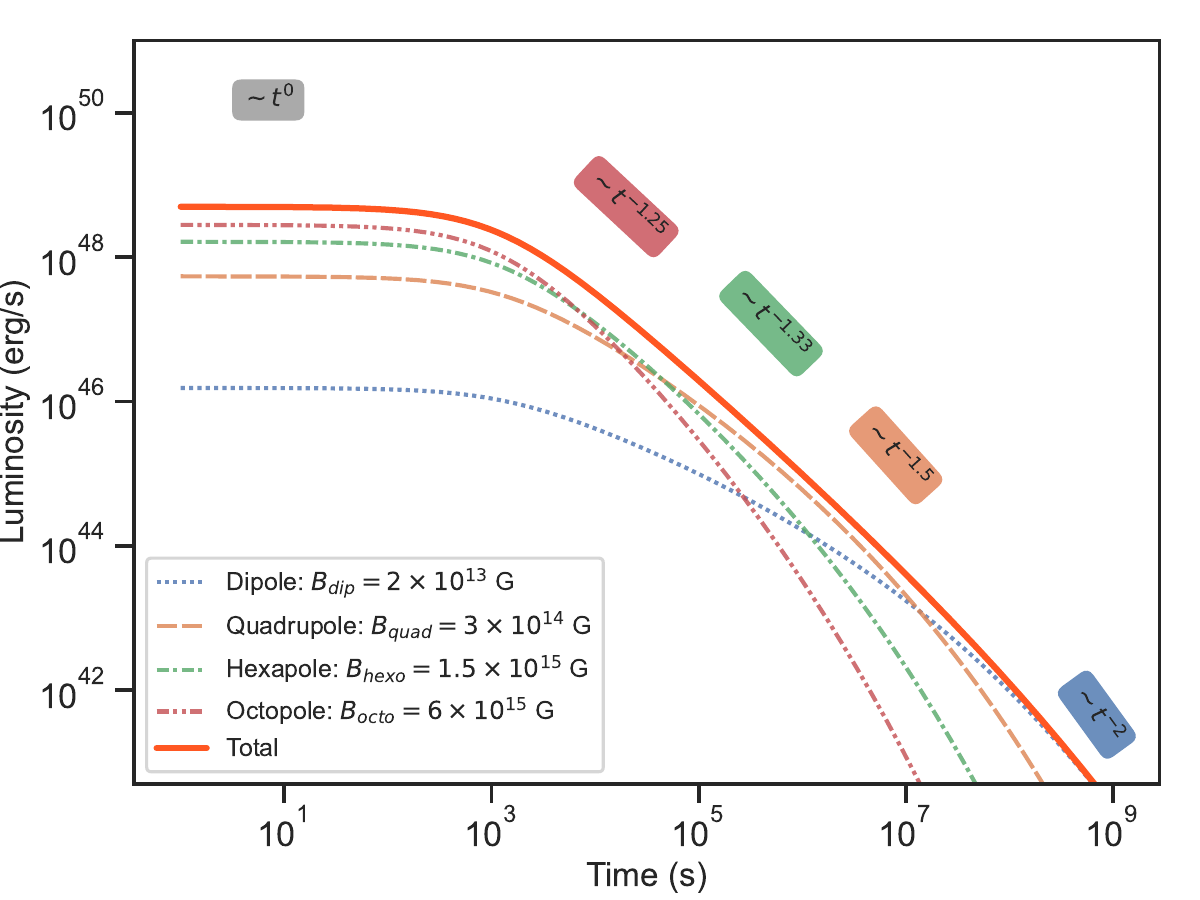}
\caption{\textcolor{black}{Spin–down luminosity contributions of a newborn magnetar endowed with multipolar fields: dipole ($B_{\rm dip}=2\times10^{13}\,\mathrm{G}$), quadrupole ($B_{\rm quad}=3\times10^{14}\,\mathrm{G}$), hexapole ($B_{\rm hexa}=1.5\times10^{15}\,\mathrm{G}$), and octopole ($B_{\rm octo}=6\times10^{15}\,\mathrm{G}$).  The higher‐order multipoles, owing to their stronger field strengths, dominate the early energy release; beyond $t\simeq10^{8}\,$s, the dipole term becomes the dominant contributor to the spin–down luminosity.}}
\label{fig:multipole-example}
\end{figure}

{\color{black}

A rotating neutron star (NS) in vacuum can be decomposed into vector spherical harmonics of order $l$ (dipole $l=1$, quadrupole $l=2$, hexapole $l=3$, etc.; \citealt{barrera1985,1998clel.book.....J,ptri2015,Wang:2024fun}). Each multipole $l$ contributes its own radiative spin–down luminosity, which for order $l$ scales as
\begin{equation}\label{eq:L_l_rest}
L_l(t) \;=\; C_l\,\Omega^{2l+2}\,B_l^2\,R^{2l+4}\,\Theta_l^2
\ \,,
\end{equation}
where $\Omega$ is the NS angular velocity, $B_l$ its characteristic surface field strength, $R$ the stellar radius, $\Theta_l$ a geometric inclination factor, and $C_l$ a dimensionless constant fixed by the vacuum solution. The total spin–down luminosity is then simply the sum of all multipolar contributions:
\begin{equation}\label{eq:L_l_rest}
L_{\rm tot}(t) = \sum_{l=1}^\infty L_l(t).
\end{equation}

\noindent Since the rotational energy loss satisfies

\begin{equation}
\dot E_{\rm rot}
=I\,\Omega\,\dot\Omega
=-\,L_{\rm tot}(t)
\label{eq:energy}
\end{equation}
one immediately gets
\[I\,\Omega\,\dot\Omega
=-\sum_{l=1}^\infty C_l\,B_l^2\,R^{2l+4}\,\Theta_l^2\,\Omega^{2l+2}
\quad\Longrightarrow\quad\]
\[\dot\Omega 
=-\sum_{l=1}^\infty K_l\,\Omega^{2l+1}\,,\]
with $K_l=C_lB_l^2R^{2l+4}\Theta_l^2/I$. We treat it as a separable ordinary differential equation:


\begin{align}
\frac{d\Omega}{\displaystyle\sum_{l=1}^\infty K_l\,\Omega^{2l+1}} \nonumber
&= -\,dt, \\[1ex] 
\int_{\Omega_0}^{\Omega(t)} \nonumber
\frac{d\Omega'}{\displaystyle\sum_{l=1}^\infty K_l\,{\Omega'}^{2l+1}}
&= -\,\int_{t_0}^{t} dt'
\;=\;\,-(t - t_0).
\nonumber
\end{align}

\noindent This integral is the \emph{implicit} solution:
\begin{equation}
t \;=\;
t_0 + \int_{\Omega(t)}^{\Omega_0}
\frac{d\Omega'}{\displaystyle\sum_{l=1}^\infty K_l\,{\Omega'}^{2l+1}}.
\end{equation}
Once $\Omega$ is solved from the above equation, $L_{\rm tot}(t)$ can be obtained from Equation \ref{eq:energy}. A comprehensive treatment of the various configurations of this solution is presented in Figures 1–3 of \citet{Wang:2024fun}.   In the case of a single dominant multipole, we recover: 
\begin{align}
\int \Omega'^{-(2l+1)} \,d\Omega'
&= -\frac{1}{2l}\,\Omega'^{-2l} 
\quad\Longrightarrow\quad \nonumber \\
&\Omega(t) = \Omega_{0} \Bigl(1 + \frac{t}{\tau_l}\Bigr)^{-1/(2l)}. 
\end{align}

\noindent Substituting this result into the luminosity gives
\[
L_l(t) = L_{l,0}\bigl(1 + \tfrac{t}{\tau_l}\bigr)^{-(1 + 1/l)},
\]
where we have defined
\[
\tau_l = \frac{I\,c^{2l+1}}{(2l+2)\,C_l\,B_l^2\,R^{2l+4}\,\Theta_l^2\,\Omega_{0}^{2l}},
\quad
L_{l,0} = \frac{I\,\Omega_{0}^2}{2l\,\tau_l}.
\]

\subsection{Power-Law Approximation for Separated Multipoles}

If each multipole spins down on a distinct timescale, they act nearly independently. The approximate total luminosity is
\begin{equation}
    L_{\rm tot}(t) \approx \sum_{l=1}^\infty L_{l,0}\bigl(1 + \tfrac{t}{\tau_l}\bigr)^{-(1 + 1/l)}. \label{eq:ls} 
\end{equation}

Substituting this into the energy equation (Eq. \ref{eq:energy}) gives
\begin{equation}
\Omega(t) \approx \sqrt{\Omega_0^2 - \frac{2}{I} \sum_{l=1}^\infty l\, L_{l,0} \tau_l\bigl[1 - (1 + \tfrac{t}{\tau_l})^{-1/l}\bigr]}.
\end{equation}
which in the case of a single dominant order $l$, as $\Omega^2_{0} \simeq \frac{2}{I}l \,L_{l,0}\,\tau_{l,0}$, the well‐known closed‐form solution \citep{Wang:2024fun}
\begin{equation}
\Omega(t) \simeq \Omega_0\Bigl(1 + \tfrac{t}{\tau_l}\Bigr)^{-1/(2l)} \nonumber
\end{equation}
is recovered. This sequential multipole picture is realized when the spin-down times of successive multipoles are well separated (i.e.\ $\tau_{l+1}/\tau_{l}\ll1$), the multipolar field strengths follow a strong hierarchy ($B_{l}\,R^{\,l+2}\,\Theta_{l}\gg B_{l-1}\,R^{\,l+1}\,\Theta_{l-1}$), and the observational window lies between these timescales ($\tau_{l+1}\ll t_{\rm obs}\ll\tau_{l}$).

Figure.~\ref{fig:multipole-example} shows the spin–down luminosity contributions of a newborn magnetar endowed with multipolar surface fields—dipole ($B_{\rm dip}=2\times10^{13}\,\mathrm{G}$), quadrupole ($B_{\rm quad}=3\times10^{14}\,\mathrm{G}$), hexapole ($B_{\rm hexa}=1.5\times10^{15}\,\mathrm{G}$), and octopole ($B_{\rm octo}=6\times10^{15}\,\mathrm{G}$)—for an initial spin period of 1 ms, $R=1.0\times10^{6}$ cm, $M=1.4\,M_{\odot}$, and angular factors $\Theta_{\rm dip}^2=1$, $\Theta_{\rm quad}^2=10$, $\Theta_{\rm hexa}^2=15338$, and $\Theta_{\rm octo}^2=1179671$.  As expected, the higher‐order multipoles dominate the early energy release, but near $t\simeq10^{8}\,$s the dipole component overtakes them, marking the transition to a dipole‐dominated spin–down. 


} 

\textcolor{black}{\subsection{Vacuum vs.\ Wind Spin-Down and Magnetic Field Contributions}}

The dipolar spin-down approximation is widely employed in studies of millisecond magnetars as central engines for gamma-ray bursts (GRBs) \citep{ 2001ApJ...552L..35Z, 2010MNRAS.402..705L, 2011A&A...526A.121D, 2011AIPC.1358..195R, 2012IAUS..279..297O, 2012Sci...338.1445N, 2013MNRAS.430.1061R, 2014ApJ...785...74L, 2014MNRAS.443.1779R, 2015ApJ...805...89L, stratta2018magnetar}. However, the idea that pulsar dipolar fields prevail over multipolar radiation in newborn millisecond magnetars remains both unverified by observations and difficult to reconcile with theoretical predictions.

\textcolor{black}{For a newly born millisecond magnetar, it is plausible that higher-order multipolar fields (\(\ell>1\)) dominate the early spin-down phase (\(t\lesssim10^6\) s), since multipolar electromagnetic losses scale as \(L_\ell\propto B_\ell^2\,\Omega^{2\ell+2}\,R^{2\ell+4}\). When the light-cylinder radius \(R_{\rm LC}=c/\Omega \approx50\) km is sufficiently compact, these losses can exceed dipole wind braking because closed multipolar field lines suppress open magnetic flux and reduce wind efficiency \citep{2002MNRAS.334..743A,ptri2015,Wang:2024fun}. As the magnetar continues to spin down, its light‐cylinder radius \(R_{\rm LC}\) expands, weakening multipole torques and reopening closed field lines so that the dipole term (\(L\propto\Omega^4\)) gradually prevails.  By \(t\sim10^7\!-\!10^8\) s the classic dipole regime (\(L\propto t^{-2}\), \(\Omega\propto t^{-1/2}\)) is established, in agreement with the typical decay slopes of some GRB afterglows.  }

\textcolor{black}{In contrast, in vacuum multipole models, each \(\ell\)-pole torque vanishes for an aligned rotator (\(\alpha=0\)), since \(L_\ell\propto\sin^2\alpha\). However, in realistic plasma-filled (force-free or MHD) magnetospheres, even aligned dipoles lose rotational energy via a steady wind of luminosity \(L_{\rm wind}\sim B_1^2R^6\Omega^4(1+\sin^2\alpha)/c^3\) \citep{2006ApJ...643.1139C,1999ApJ...525L.125H}. Furthermore, plasma loading and pair cascades can flatten the pure vacuum scaling \(L_\ell\propto\Omega^{2\ell+2}\); MHD simulations indicate that while \(\Omega(t)\propto t^{-1/(2\ell)}\) remains a useful approximation, the actual braking indices are shifted by plasma effects \citep{2006ApJ...648L..51S,2015ApJ...801L..19P}.}

 \textcolor{black}{A newborn, millisecond‐period magnetar may lose its rotational energy predominantly via a Poynting‐flux–dominated wind.  In the canonical magnetic‐dipole spin‐down scenario, continuous injection of this wind into the external blast wave—analogous to refreshed‐shock models—can account for the shallow X‐ray plateau, and, under favorable conditions, the coeval optical plateau observed in many GRB afterglows, provided that the magnetic energy is efficiently converted into bulk kinetic energy \citep{2011A&A...526A.121D,2001ApJ...552L..35Z}.  Models incorporating higher‐order multipolar components demonstrate that quadrupole and octupole terms may dominate the Poynting output close to the magnetar surface \citep{ptri2015}; however, a self‐consistent treatment of how a multipolar Poynting wind couples to the forward shock has yet to be developed.  }

\textcolor{black}{Although multipolar spin-down models can qualitatively reproduce many observed GRB plateau and decay indices, persistent early multipolar dominance remains speculative. Definitive validation will require high-cadence X-ray and optical afterglow observations alongside advanced MHD simulations that incorporate plasma inertia, toroidal fields, and realistic multipolar geometries. Moreover, how plasma effects modify multipole-driven spin-down in newborn magnetars is still an open question.}

\section{The Dainotti Relation whithin the Magnetar Multipolar Fields}\label{sec:Dai}

\subsection{Approaches to Fitting Multipolar Spin-Down}

\textcolor{black}{\paragraph{Full Multipolar Fit} In principle, one could treat \(\{\tau_\ell,\,L_{\ell,0}\}_{\ell=1}^L\) as free parameters up to some truncation \(\ell_{\max}\).  The complete fit would then simultaneously model all ``breaks'' at \(t\sim\tau_\ell\).  However, this introduces \(2\,\ell_{\max}\) parameters (plus any geometric factors \(\Theta_\ell\)), leading to strong degeneracies. Even high-quality afterglow data rarely constrain more than two breaks, so a simultaneous multi-\(\ell\) fit is both underdetermined and prone to overfitting. Once \(t\gg\tau_\ell\), \(L_\ell(t)\propto t^{-1-1/\ell}\) decays steeper than any lower order.  Thus after a transient early phase, a single \(\ell\) will always dominate.  By the time a higher-\(\ell\) term has steepened sufficiently, it is already subdominant.}

\textcolor{black}{\paragraph{Hierarchical ``Mixed-\(\ell\)'' Fit} One can impose a prior ordering \(\tau_1>\tau_2>\cdots>\tau_{\ell_{\max}}\), reflecting that higher-\(\ell\) terms decay faster.  Then proceed stepwise: (i) fit the earliest break to determine \(\tau_{\ell_{\max}}\), (ii) remove that component’s contribution, (iii) fit the next break, and so on.  This hierarchical approach reduces degeneracy but still requires detecting each individual break.  In practice, intermediate breaks are shallow and often missed.}

\textcolor{black}{\paragraph{Single-\(\ell\) Fit as Leading Approximation} Instead, one identifies the single dominant order \(\ell_{\rm dom}\) whose \(\tau_\ell\) lies within the observed plateau duration.  One then fits
\[
L(t)\approx L_{\ell_{\rm dom},0}\Bigl(1 + \tfrac{t}{\tau_{\ell_{\rm dom}}}\Bigr)^{-1-1/\ell_{\rm dom}},
\]
This introduces just two parameters \((L_{\ell_{\rm dom},0},\,\tau_{\ell_{\rm dom}})\), plus an implied \(\ell_{\rm dom}\).  An F-test then confirms whether adding \textcolor{black}{additional multipoles} significantly improves the fit; typically, it does not \citep{Wang:2024fun}. From a physical standpoint, the observed X-ray afterglows \textcolor{black}{may be} reproduced by a magnetar endowed with a multipolar field—dipole, quadrupole, hexapole, octopole, and higher orders—rather than by a pure dipole alone. In this picture, the multipole whose surface field is strongest ($\ell_{\rm dom}>1$) sets the characteristic spin-down timescale $\tau_{\ell_{\rm dom}}$, and therefore governs the early plateau and first break in the light curve. As that dominant higher-order component rapidly decays, the canonical dipole ($\ell=1$)—with its longer spin-down time—inevitably takes over at later times, producing the familiar late-time $t^{-2}$ decline (see, e.g., Figure~3 of \citealt{Wang:2024fun}). }

 \textcolor{black}{Because of these factors, a \emph{single-\(\ell\) fit} captures the leading spin-down and luminosity evolution over the observed interval without overfitting.  Consequently, one obtains reliable estimates of \(\tau_\ell\) and \(L_{\ell,0}\), which in turn yield the Dainotti slope. As previously shown,} the spin-down time scale, $\tau_l$ and the plateau luminosity, $L_{l,0}$ of a magnetar are related as:  
 
\begin{align}
    L_{l,0} = \frac{I \Omega_0^2}{2l \tau_l}, \nonumber \\
    L_{l,0} \propto \tau_l^{-1}.
\label{eq:LT}
\end{align} 

The Dainotti relation, given by $\log L_X = \log a + b \log T_a^*$, links the isotropic plateau luminosity $L_X$ to the rest-frame plateau end time $T_a^*$ in GRB X-ray afterglows. Observationally, the intrinsic slope $b$ is measured as $b = -1.07_{-0.14}^{+0.09}$ \citep{2008MNRAS.391L..79D, 2010ApJ...722L.215D, 2011ApJ...730..135D, 2011MNRAS.418.2202D, 2013ApJ...774..157D}, consistent with the magnetar spin-down luminosity model, where $L_X$ corresponds to $L_{l,0}$ and $T_a^*$ to $\tau_l$.

Crucially, the dependence of $\tau_l$ on the plateau properties is independent of the multipolar order $l$, emphasizing the universality of the Dainotti relation within the multipolar magnetic field frameworks.

\begin{figure}
\centering
\includegraphics[width=1\hsize,clip,angle=0]{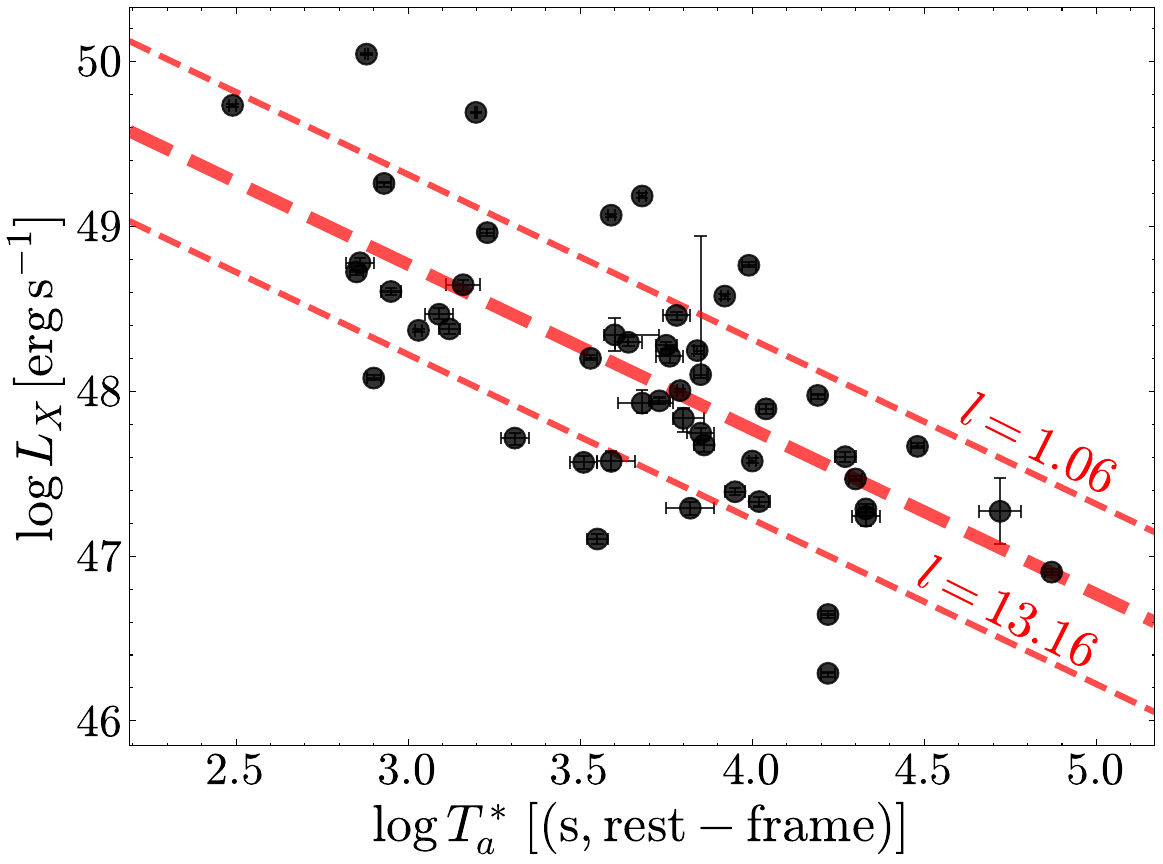}
\caption{\textcolor{black}{Platinum Sample of GRBs taken from \cite{Cao2022}. The upper and lower thin dashed lines denote the 1 $\sigma$ dispersion lines of the data, with an upper envelope corresponding to $l=1.06$ ($ L_X = 2.07 \times 10^{52}\,T_a^{*-1} $), the lower envelope corresponding to $l= 13.16$ ($ L_X = 1.67 \times 10^{51}\,T_a^{*-1} $), and the mean fit line corresponds to $l=3.74$ assuming a magnetar with an initial spin period of 1 ms. }}
\label{fig:dainotti-ul}
\end{figure}

Multipolar configurations reproduce the Dainotti relation, irrespective of the dominant magnetic component:

\begin{itemize}
    \item The observed slope $b \approx -1$ holds across all multipolar orders, while the model accommodates decay indices ranging from $-2$ to $-1$ in Swift-XRT GRB afterglows, highlighting the model's capability to describe the diverse features of GRB X-ray plateaus.

    \item \textcolor{black}{When modeling the electromagnetic emission from a newborn NS, a spin period of $1$ ms is commonly adopted as the shortest stable value \citep{2007Ap&SS.308..119D, 2016PhRvD..93d4065G}. The main physical reason lies in the break-up limit: if the star spins faster than approximately $1$ ms, the centrifugal force exceeds its own gravity, leading to mass shedding \citep{1983bhwd.book.....S}. A rough derivation can be obtained by equating the gravitational and centrifugal forces:
    \begin{equation}
    P_{\min}\simeq 0.54 \left({M\over M_\odot}\right)^{-1/2}
    \left({R\over10{\rm~km}}\right)^{3/2}{\rm~ms} ,
    \end{equation}
    Taking into account relativistic calculations and realistic equations of state (EOS), \citet{2004Sci...304..536L, 2007PhR...442..109L} derived a more accurate minimum spin period:
    \begin{equation}
    P_{\min}\simeq 0.96 \left({M\over M_\odot}\right)^{-1/2}
    \left({R\over10{\rm~km}}\right)^{3/2}{\rm~ms} .
    \end{equation}
    For a typical NS with mass $M = 1.4 M_\odot$ and radius $R = 10$ km, this refined equation gives a minimum spin period of $P_{\min} = 0.8$ ms. If changing the radius $R$ to $12$ km, then the minimum spin period of $P_{\min} = 1$ ms. Furthermore, at sub-millisecond spin periods, gravitational-wave emission triggered by r-mode or other instabilities can rapidly remove angular momentum \citep{1998PhRvD..58h4020O, 1998ApJ...502..708A}, causing fast spin-down. In practice, no NS has been observed to spin faster than $1.4$ ms \citep{2006Sci...311.1901H}. Consequently, a spin period of $1$ ms is typically adopted as the upper limit in electromagnetic energy injection models for gamma-ray bursts, as well as in this article to compute the upper boundary of the spin-down luminosity. Therefore, for fiducial values of $M = 1.4\, M_\odot$, $R = 10\, \mathrm{km}$, and $P_0 = 1\, \mathrm{ms}$, using Eq. \ref{eq:LT}, we can determine the upper limit (UL) of $L_{l,0}$ as
    \begin{equation}
    L_{l,0}^{\mathrm{UL}} = \frac{2.2 \times 10^{52}}{l~ \tau_l}.
    \label{eq:upperlimit}
    \end{equation}
    This upper limit of the Dainotti relation reflects the maximum spin-down luminosity that a newly formed magnetar produces during the plateau phase of a GRB.
     \textcolor{black}{We utilize the platinum sample of 50 GRBs by \cite{Cao2022}} as shown in Figure \ref{fig:dainotti-ul}}. \textcolor{black}{We fit for the mean and $1 \sigma$ dispersion lines normalizations; from the mean fit, we find $l=3.74$, corresponding to}
    \begin{equation}
  L_{l,0}^{\mathrm{UL,obs}}
    = \frac{5.87 \times 10^{51}}{\tau_l}\,.
\end{equation}
\textcolor{black}{Likewise, the upper dispersion envelope of $ L_X = 2.07 \times 10^{52}\,T_a^{*-1} $ yields $l=1.06$, while the lower envelope  ($ L_X = 1.67 \times 10^{51}\,T_a^{*-1} $) yields $l=13.16$. Since both of these best‐fit values exceed unity, they reinforce the need for higher‐order magnetic moments beyond a pure dipole.  In other words, within the framework of magnetar spindown, allowing for multipolar contributions provides a far better match to the observed normalization of the Dainotti relation than a simple $l=1$ spin‐down law. }

One might argue that reducing the initial spin period of the magnetar could allow dipole radiation to account for $L_X T^*_a \lesssim 10^{52}$. However, considering the 3D version of the Dainotti relation \citep{2016ApJ...825L..20D}, which correlates $T^*_a$, $L_X$, and the peak luminosity of the prompt emission ($L_{\rm peak}$), $L_X T^*_a \sim 10^{52}$ corresponds to $L_{\rm peak} \sim 10^{53}~\rm erg/s$. According to \cite{Beniamini:2017ilu}, achieving such high peak luminosities ($L_{\rm peak} \sim 10^{53}~\rm erg/s$) requires an initial spin period $P_0 \sim 1~\rm ms$. For periods greater than $1-1.5~\rm ms$, the maximum isotropic luminosity of the magnetar falls short of $10^{53}~\rm erg/s$. Therefore, dipole radiation primarily accounts for the upper limit of the plateau energy in this context.

\item For any multipole order, the plateau luminosity and duration scale with $\tau_l$, ensuring consistent spin-down behavior across magnetic configurations. During the early phases, GRB afterglows may arise from a combination of multipolar fields. Higher-order multipoles typically dominate the spin-down initially, although slight deviations from a strict power-law decay with an index of $-1$ may occur due to the coexistence of multiple field components \citep{Wang:2024fun}. 

\end{itemize}

Evolutionary changes in the magnetic field structures, including the emergence or decay of multipolar components, may introduce variability in power-law indices and luminosity evolution \citep{2021Univ....7..351I}. These changes, contribute to observational scatter without undermining the the Dainotti relation.

The interplay between dipolar and multipolar magnetic fields offers a coherent framework for understanding GRB afterglows and the Dainotti relation. Although early GRB phases may exhibit deviations due to multipolar dominance, the underlying connection between spin-down luminosity and plateau properties remains robust.

\subsection{Impact of Radiative Efficiency and Jet Opening Angle on the Dainotti Relation}\label{sec:eff}

\textcolor{black}{We have thus far assumed 100\% radiative efficiency and isotropic emission across all GRBs, which is an unphysical assumption. Moreover, a bolometric correction factor $k$ should be included. The final true luminosity is then}

\begin{equation}
\textcolor{black}{L_{X,\text{true}} = \frac{\theta_{\rm j}^2}{2}  \frac{ L_{\text{sd}}}{\epsilon_X} \times k}
\label{eq: corr}
\end{equation}

\textcolor{black}{where $L_{\text{sd}}$ is the observed spin-down luminosity. The opening angle and structure of GRB afterglows is still under debate, but typically differ between SGRBs (Type-I) and LGRBs (Type-II). For SGRBs, we take a median jet opening angle $\theta_j = 16^\circ \pm 10^{\circ}$ \citep{Fong:2015oha}. For LGRBs, $\theta_j$ is typically found to be lower, often peaking at $< 10^{\circ}$ \citep{ 2001ApJFrail, 2003ApJBloom, 2005ApJGuetta, goldstein2016estimating}. It is difficult to quote a single ``average'' efficiency, as the exact value depends on the assumed model and their paramaters. Many early afterglow modeling efforts simply assumed a constant efficiency of
order $\sim10\%$ (i.e. 0.1) for converting spin-down power into X-ray afterglow luminosity. For instance, \cite{nava2013afterglow} took a ``semiradiative'' external‐shock approach in which they fold the microphysical parameters into a single overall efficiency $\textcolor{black}{\epsilon}$ and explicitly follow its time‐dependence.  As the blast wave transitions from a fast‐cooling (fully radiative) regime to the standard adiabatic evolution,  they find that the radiative efficiency reaches a canonical late‐time value of $ \textcolor{black}{\epsilon} \simeq \textcolor{black}{0.1}
  \quad\text{by}\quad
  \textcolor{black}{t}\sim\textcolor{black}{10^2}\;\textcolor{black}{\mathrm{s}},
$.
However, more detailed theoretical work indicates the true efficiency may be lower. For example, \cite{2019ApJDai} consider a scenario where the X-ray plateau is produced by internal
dissipation in an ultrarelativistic, Poynting-flux dominated wind (a ``magnetar wind nebula'' scenario). In their fits to many GRB plateaus, the inferred X-ray efficiencies are typically below 10\%.}

\textcolor{black}{The question then becomes, if both opening angle and radiative efficiency are accounted for in the entire population, what would be the net effect of the normalization and dispersion of the data?
Given the above information, we adopt fiducial average values of $\theta_j = 10^{\circ}$ and $\epsilon = 0.1$. For bolometric correction, typical factors of $k$ range from 0.4 to 7 \citep{Bloom:2001ts, 2013MNRAS.430.1061R}, yielding a fiducial average of 3.7. With these fiducial values, Eq. \ref{eq: corr} would yield a factor of $0.56 \times L_{\rm sd} $. If a lower radiative efficiency of $\epsilon =0.05$ were assumed, in alignment with \cite{2019ApJDai}, the correction factor would be on the order of unity. Hence, on a population level, it is reasonable to assume that the normalization would not change significantly if $\theta_j$ and $\epsilon$ were known for all GRBs. It is more difficult to infer the effect on dispersion; however one can reasonably assume a tighter Dainotti correlation when accounting for $\theta_j$.  A similar reduction in dispersion was demonstrated by \cite{Ghirlanda2004}, where they introduce a collimation correction for the 
Amati relation, $(1-\cos\theta_j)\times E_{\rm iso} - E_{\rm peak}$. A tighter correlation would compress our inferred multipole‐order range—currently $1.06 < l < 13.16$ (shown in figure \ref{fig:dainotti-ul})— presumably to a narrower, and therefore more physically plausible, interval centering around $l \sim 3.7$. One would expect a similar tightening when correcting for efficiency.} 

\textcolor{black}{Moreover, \citet{2014MNRAS.443.1779R} showed that if one assumes 100\% spin-down energy conversion efficiency and isotropic emission across GRBs, magnetar central engines reproduce the observed Dainotti correlation within its uncertainties. They further explored parameter variations—adopting a fixed jet half–opening angle of $1^\circ$ and efficiencies between 25\% and 99\%; as previously adopted by \citep{2012A&A...539A...3B}—and found that the resulting normalization range exceeds the observed dispersion. Clearly, different combinations of opening angle and efficiency can shift this normalization significantly, and variations in these factors among GRBs contribute to the scatter of the correlation. Consequently, while their work supports the magnetar explanation under idealized assumptions, it also highlights that the inferred Dainotti scatter is sensitive to model-dependent angular and efficiency factors.   Similarly, \citet{2015ApJ...800...31D} emphasized that selection effects and unknown efficiency functions can influence the evaluation of the intrinsic slope of the Dainotti relation. Therefore, attributing the plateau emission to internal dissipation of the magnetar central engine necessitates assuming constant efficiency and geometric parameters, which may not always be exact. }

\section{Discussion and Conclusions}\label{sec:conclusion}

The Dainotti relation has long provided an empirical framework for studying the physical mechanisms governing GRB afterglows, but its origins have remained unclear within traditional models. This study addresses this gap by presenting a \textcolor{black}{plausible} explanation rooted in the multipolar magnetic field structures of newborn magnetars. Unlike the conventional dipole approximation, which struggles to account for the diversity in GRB afterglow behaviors, incorporating higher-order multipoles offers a more comprehensive understanding of energy release and spin-down dynamics.

An extensive analysis of the Swift-XRT database reveals consistent patterns in GRB plateau decay indices, supporting the significant role of multipolar fields in shaping afterglow phases. The derived spin-down luminosity scaling, applicable across all multipolar orders, provides a robust theoretical foundation for the Dainotti relation. This scaling elucidates the connection between plateau luminosity, duration, and the magnetar's spin-down timescale, offering deeper insights into the central engines of GRBs.

Higher-order multipolar magnetic fields, unlike simple dipole fields, facilitate more complex energy dissipation processes through localized regions of intense field strength and intricate geometries. These configurations introduce additional degrees of freedom in energy extraction and radiation, resulting in variable timescales and luminosities that \textcolor{black}{can} reproduce the Dainotti correlation. By incorporating non-dipolar components such as quadrupole and octupole moments, multipolar fields better align with the observed diversity of GRB afterglows, bridging theoretical predictions and empirical data \citep{10.1093/mnras/stt872,2017A&A...600A..98D,2020ApJ...893..148R,Wang:2024fun}.

The analytical solutions for multipolar electromagnetic fields and spindown luminosities presented in section \ref{sec:model} assume the NS to be in a vacuum. For non-vacuum environments, numerical simulations by \citep{spitkovsky2006time} have shown that the surrounding plasma would work to decrease the braking index, $n$ (where $\frac{d\Omega}{dt} = -K\Omega^{n}$), which would in turn increase the afterglow power law index. This shows that, in fact, a more comprehensive treatment would mean that even afterglows well fitted with dipolar models are likely to have significant contributions from higher orders. 

\textcolor{black}{Despite this promising result, several key assumptions limit the robustness of our conclusions.  Our interpretation rests on the premise that the observed plateau emission is powered by the magnetar wind/magnetic field.  Alternative scenarios—most notably continuous energy injection into the external shock—can produce plateau features without invoking multipolar fields, potentially rendering multipolar spin‐down unnecessary in some cases.  Third, our sample remains subject to observational selection biases and uncertainties in bolometric corrections, especially given the reliance on Swift‐XRT data (0.3--10 keV) and an assumed broadband scaling factor.}

\textcolor{black}{In addition, it has been suggested that black‐hole spin‐down—via the Blandford–Znajek process operating in a magnetically arrested disk (MAD)—can likewise power the X-ray plateau, providing an alternative central‐engine scenario that reproduces an $L_X\propto T_a^{*-1}$ scaling \citep{Lenart2025}. This cannot be ruled out, and indeed it is likely that in reality, some GRB afterglows are powered by black hole spindown, and others by magnetar spindown. Our aim here is to show that within the framework of magnetar spindown, a pure dipolar contribution is likely insufficient. A comprehensive, statistical analysis comparing the prevalence of magnetar versus black‐hole spin‐down signatures in GRB X-ray plateaus would be a valuable direction for future work. }

\textcolor{black}{Given these caveats, we emphasize that while the multipolar spin‐down scenario offers a self‐consistent explanation for many observed X‐ray plateaus, the precise value of the multipole index remains uncertain. Future work should aim to constrain radiative efficiencies and jet opening angles on a burst‐by‐burst basis, incorporate contemporaneous multiwavelength afterglow data to reduce bolometric uncertainties, and explore hybrid models in which both internal magnetar dissipation and external shock processes/black hole spindown contribute to the plateau phase.  Only through such efforts can we definitively determine the true magnetic geometry and spin‐down behavior of the central engine in GRBs.  }

\bigskip

\acknowledgments

 \textcolor{black}{We would like thank the anonymous referee for her/his insightful comments and constructive suggestions, which have significantly strengthened the quality, clarity, and the presentation of this work.} ESY acknowledges support from the ``Alliance of International Science Organization (ANSO) Scholarship For Young Talents”. R. Moradi acknowledges support from the Institute of High Energy Physics, Chinese Academy of Sciences (E32984U810) and the Beijing Natural Science Foundation (IS24021). 


\end{document}